\documentclass{article}
\usepackage{arxiv}
\usepackage[utf8]{inputenc} 
\usepackage[T1]{fontenc}    
\usepackage{hyperref}       
\usepackage{url}            
\usepackage{booktabs}       
\usepackage{amsfonts}       
\usepackage{nicefrac}       
\usepackage{microtype}      
\usepackage{graphicx}
\usepackage{doi}
\usepackage[english]{babel}
\usepackage{amsmath, amssymb}
\usepackage{mathtools}
\usepackage{pgf,tikz,svg}
\usepackage{forloop}
\usepackage{ifthen}
\usepackage{hyperref}
\usepackage{csquotes}
\usepackage[onehalfspacing]{setspace}

\title{Modeling of a multiple source heating plate}

\date{\today} 					

\author{ \href{https://orcid.org/0000-0003-1762-5659}{\includegraphics[scale=0.06]{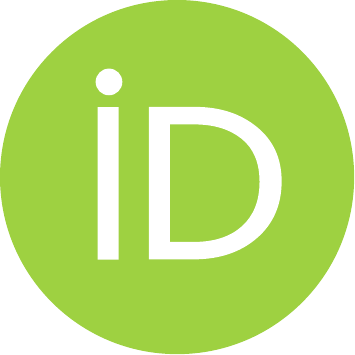}\hspace{1mm}Stephan Scholz}, Lothar Berger \\
	Control and Process Engineering\thanks{Web: \href{https://www.rwu.de/en/research/labs/control-and-process-engineering}{https://www.rwu.de/en/research/labs/control-and-process-engineering}}\\
	RWU Ravensburg-Weingarten University of Applied Sciences\\
	Weingarten, Germany\\
	Email to: \texttt{stephan.scholz@rwu.de} \\
}

\renewcommand{\shorttitle}{Modeling of a multiple source heating plate}

\hypersetup{
pdftitle={Modeling of a multiple source heating plate},
pdfsubject={cs.CE, cs.SY},
pdfauthor={Stephan Scholz, Lothar Berger},
pdfkeywords={Multiple Sources, Heating Plate, Heat Equation},
}

\begin{document}
\maketitle

\begin{abstract}
	
Heating plates describe the transfer of heat from actuators to a target object. In other words, they separate the heat sources and heated object and can be further used to apply a specific heat distribution on this object. Therefore, an exact description of their thermal dynamics and an efficient coordination of their actuators is necessary to achieve a desired time-dependent temperature profile accurately.
In this contribution, the thermal dynamics of a multiple source heating plate is modeled as a quasi-linear heat equation and the configuration of the spatially distributed actuators and sensors are discussed. Furthermore, the distributed parameter system is approximated using a Finite Volume scheme, and the influence of the actuators' spatial characterization on the plate's thermal dynamics is studied with the resulting high-dimensional system.

\end{abstract}


\newcommand{\heat}{\vartheta}
\newcommand{\heattx}{\heat(t,x)}
\newcommand{\heattxi}{\heat(t,\xi)}
\newcommand{\dheattx}{\dot{\heat}(t,x)}
\newcommand{\temperature}{\theta}
\newcommand{\opd}{\operatorname{d}}
\newcommand{\cyl}{Q}
\newcommand{\linheattx}{w(t,x)}
\newcommand{\appheattx}{g(\linheattx)}
\section*{Introduction}

Thermal processing for advanced manufacturing, like semiconductor fabrication \cite{book:xiao2012introduction} , additive manufacturing \cite{article:wei2019heat}, or biotechnology processing \cite{article:zhou2011optimizing}, requires ever higher precise temperature controllability, temperature ramp-up performance, and temperature profile stability. This is achieved by thermal systems with multiple spatially distributed heating sources, and cooling, for example, multiple source heating plates \cite{article:berger2007qualification}.

Since existing physical models are insufficient to achieve the required precision, existing control schemes for multiple source heating plates depend on data-driven models \cite{article:berger2004global} or various lumped models \cite{article:tay2008predictive}-\cite{article:jatunitanon2018robust}. 


Existing control schemes for multiple source heating plates also utilize offline calibration methods like wireless temperature sensor arrays \cite{article:berger2004global} because no observability models for multiple spatially distributed heating sources were known. Meanwhile, there exist observer design based approaches for one-dimensional thermal problems using boundary \cite{article:krstic2005backstepping,article:krstic2015adaptive} or in-domain control and measurement \cite{article:kharkovskaia2020interval}, however if these are applicable to multiple spatially distributed heating sources problems, requires further investigation.

Hence, renewed interest has come to physical modeling and designing efficient solvers for multiple spatially distributed heating sources thermal processing \cite{article:eppler2001optimal}. Due to ever increasing process requirements, there is also renewed interest in observability modeling \cite{chapter:seidman1996control} to eliminate the need for offline calibration methods.

Therefore, modeling and control of the heat equation \cite{article:tan2000study} and related reaction-diffusion equation \cite{article:song2016nonlinear} - and observability \cite{article:tutcuoglu2016nonlinear} - were discussed extensively in recent years.  For example, in \cite{article:utz2011trajectory} the authors approximate a two-dimensional quasi-linear heat equation using the finite difference method and apply on its resulting high-dimensional nonlinear system flatness-based methods to reach trajectory planning. The authors of \cite{article:xiao2016sliding,article:xiao2018dominant} construct a sliding-mode controller and observer for a cylindrical rapid thermal processing system which is modeled as a nonlinear one-dimensional heat equation. 

In this work, the thermal dynamics of a cuboid heating plate is introduced as a three-dimensional distributed parameter system in the next section with boundary conditions. The idea of actuator- and sensor-typical spatial characterization is introduced in section two and a numerical two-dimensional example which illustrates the influence of the actuators' spatial characterization is discussed from section three to five. 

This work states first ideas to model and control the quasi-linear heat equation in two (and three) dimensions with actuators and sensors limited by their spatial characterization. This publication neither states a complete analytical or numerical treatment of the quasi-linear heat equation, nor yet an advanced controller design.
\section{Problem formulation}

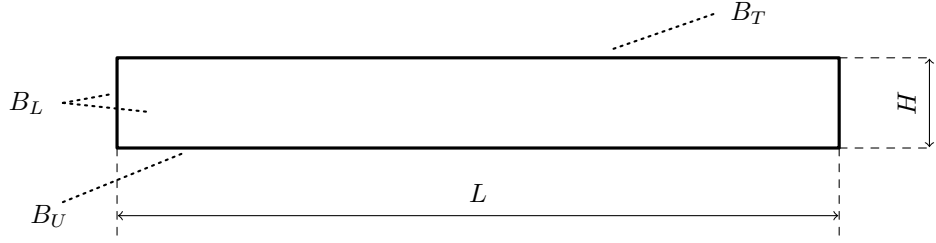
\begin{figure}[t]
		\centering
		\begin{tikzpicture}[scale=0.6,line cap=round,line join=round,x=1.0cm,y=1.0cm]
		\draw[line width=1.2] (0.,0.) -- (16.,0.) -- (16., 2.) -- (0., 2.) -- cycle;
		\draw [dashed] (0., 0.) -- (0., -2.);
		\draw [dashed] (16., 0.) -- (16., -2.);
		\draw [dashed] (16., 0.) -- (18., 0.);
		\draw [dashed] (16., 2.) -- (18., 2.);
		\draw [<->] (0.02, -1.5) -- (15.98, -1.5);
		\draw [<->] (18.0, 0.02) -- (18.0, 1.98);
		\draw [dotted, line width = 0.8] (-1.2, 1.) -- (-0.1, 1.2);
		\draw [dotted, line width = 0.8] (-1.2, 1.) -- ( 0.7, 0.8);
		\draw [dotted, line width = 0.8] (-1.2, -1.2) -- (1.5, -0.1);
		\draw [dotted, line width = 0.8] (11, 2.2) -- (13.3, 3.0);
		\draw[color=black] (8., -1.) node {$L$};
		\draw[color=black] (17.5, 1.) node {\rotatebox{90}{$H$}};
		\draw[color=black] (14., 3.) node {$B_{T}$};
		\draw[color=black] (-2., 1.) node {$B_{L}$};	
		\draw[color=black] (-1.5, -1.5) node {$B_{U}$};	
		\end{tikzpicture}
		\caption{Front view on heating plate with length L and height H. Boundaries \(B_{U}\), \(B_{T}\) and \(B_{L}\) denote the underside, topside and lateral surfaces. Actuators are applied on the underside and sensors are assumed on the topside.}
		\label{fig:plate_front_view}
	\end{figure}

The heating plate is assumed as an open cuboid
\begin{align*}
	\Omega := \left(0, L\right) \times \left(0,W\right) \times\left(0, H\right) \subset \mathbb{R}^{3}
\end{align*}
with a constant length \(L > 0\), width \(W > 0\) and height \(H > 0\).  The cuboids surface \(\partial \Omega := B_{U} \cup B_{T} \cup B_{L} \) is subdivided in the disjoint areas \(B_{U}\) on the underside, \(B_{T}\) on the topside and \(B_{L}\) on the lateral surfaces (see Figure \ref{fig:plate_front_view}).
The cuboid's temperature \(\heat: \left[0, T_{final}\right] \times  \overline{\Omega} \rightarrow \mathbb{R}_{\geq 0} \) in Kelvin varies in time and space depending on the induced \(\Phi_{in}\) and emitted heat flux  \(\Phi_{out}\) on \(B_{U}\) and  on \(B_{T} \cup B_{L} \), respectively.  The heating plate's physical properties 
\begin{itemize}
	\item thermal conductivity \( \lambda: \mathbb{R}_{\geq 0} \rightarrow \mathbb{R}_{\geq 0}\) in \(\left[\frac{W}{m~K}\right]\),
	\item specific heat capacity \( c: \mathbb{R}_{\geq 0} \rightarrow \mathbb{R}_{\geq 0}\) in \(\left[\frac{J}{kg~K}\right]\) and 
	\item density \( \rho: \mathbb{R}_{\geq 0} \rightarrow \mathbb{R}_{\geq 0} \) in \(\left[\frac{kg}{m^3}\right]\)
\end{itemize}
are assumed as continuous functions of the temperature.

The evolution of heat in the plate is modeled as quasi-linear heat equation
\begin{align}
\rho(\heattx) ~ c(\heattx) ~ \frac{\partial \heattx}{\partial t} ~=~ \operatorname{div} \left[ \lambda(\heattx) ~ \nabla \heattx \right] \label{eq:quasi_lin_heat}
\end{align}
for \( (t,x) \in \left(0,T_{final}\right) \times \Omega  \)  with initial value \( \heat(0,x) = \heat_{init}(x) \) for \( x \in \overline{\Omega} \) and boundary condition
\begin{align}
	\lambda(\heattx) ~\frac{\partial \heattx}{\partial x} \cdot \vec{n} ~=~ 
	\begin{cases}
		\phi_{in}(t,x) & ~ \text{on} ~ B_{U} \text{,} \\
		\phi_{out}(t,x) & ~ \text{on} ~ B_{T} \cup B_{L} \text{.}
	\end{cases}
	\label{eq:bc_in_out}
\end{align}
Vector \( \vec{n} \) denotes the normal vector on the described boundary. Flux \( \phi_{in} \) denotes the induced heat flux on the underside of the heating plate and \( \phi_{out} \) denotes the emitted heat flux on the topside and the lateral surfaces. The induced heat is stored in the plate and emitted via linear heat conduction
\begin{align*}
-h ~ \left[\heattx - \heat_{amb}(x)\right]
\end{align*}
with coefficient \(h \geq 0\) in \(\left[\frac{W}{m^2 ~ K}\right]\) and nonlinear heat radiation
\begin{align*}
-\epsilon(\heattx) \varrho  ~ \left[\heattx^{4} - \heat_{amb}(x)^{4}\right]
\end{align*}
with temperature-dependent emissivity \(\epsilon: \mathbb{R}_{\geq 0} \rightarrow \left[0,1\right]\), Stefan-Boltzmann constant \(\varrho \approx 5.67 \cdot 10^{-8} \) in \(\left[\frac{W}{m^2 ~ K^4}\right]\) and ambient temperature \(\heat_{amb}: B_{T} \cup B_{U} \rightarrow \mathbb{R}_{\geq 0} \) (see also \cite{book:baehr2013heat}). The emitted heat flux results in a sum of both terms as
\begin{align}
\phi_{out}(t,x) := -h ~ \left[\heattx - \heat_{amb}(x)\right] -  \epsilon(\heattx) \varrho ~ \left[\heattx^{4} - \heat_{amb}(x)^{4}\right] \label{eq:bc_out_conduction_radiation} 
\end{align}
for \((t,x) \in \left[0, T_{final}\right] \times B_{T} \cup B_{U}\).

\begin{table}[b]
	\begin{center}
		\caption{\MakeUppercase{Assumed physical Properties and Constants}}
		\begin{tabular}{|l lr|}
			\hline
			\textbf{Coefficient} & \textbf{Symbol} & \textbf{Value}  \\
			\hline
			Mass density & \(\rho\) 	& \(7800\) \\[1ex]
			Heat capacity & \(\left\{c_{0}, c_{1}\right\}\) & \(\left\{330, 0.4\right\}\) \\[1ex]
			Thermal conductivity & \(\left\{\lambda_{0}, \lambda_{1}\right\}\) & \(\left\{10, 0.1\right\}\) \\[1ex]
			Heat transfer & \(h\) 	& \(10\) \\[1ex]
			Emissivity & \(\varepsilon\) & \(0.6\) \\[1ex]
			Ambient temperature & \(\heat_{amb}\) & \(300\) \\[1ex]
			\hline
		\end{tabular}
		\label{table:phys_properties_const}
	\end{center}
\end{table}

\section{Spatial characterization}

\label{sec:spatial_char}

\begin{figure}[t]
		\centering
		\begin{tikzpicture}[scale=0.5,line cap=round,line join=round,x=1.0cm,y=1.0cm]
		\draw (0.,0.) -- (16.,0.) -- (16., 8.) -- (0., 8.) -- cycle;
		\draw [dashdotted] (4.,0.) -- (4.,8.);
		\draw [dashdotted] (8.,0.) -- (8.,8.);
		\draw [dashdotted] (12.,0.) -- (12.,8.);
		\draw [dashdotted] (0.,4) -- (16.,4.);
		\draw [dashed] (0., 0.) -- (0., -2.);
		\draw [dashed] (16., 0.) -- (16., -2.);
		\draw [dashed] (16., 0.) -- (18., 0.);
		\draw [dashed] (16., 8.) -- (18., 8.);
		\draw [<->] (0.02, -1.5) -- (15.98, -1.5);
		\draw [<->] (18.0, 0.02) -- (18.0, 7.98);
		\draw[color=black] (8., -0.9) node {\small $L$};
		\draw[color=black] (17.5, 4.) node {\small \rotatebox{90}{$W$}};
		\draw[color=black] (2., 2.) node {$\beta_{1}$};
		\draw[color=black] (6., 2) node { $\beta_{2}$};
		\draw[color=black] (10., 2) node { $\beta_{3}$};
		\draw[color=black] (14., 2) node { $\beta_{4}$};
		\draw[color=black] (2., 2 + 4.) node { $\beta_{5}$};
		\draw[color=black] (6., 2 + 4.) node { $\beta_{6}$};
		\draw[color=black] (10., 2 + 4.) node { $\beta_{7}$};
		\draw[color=black] (14., 2 + 4.) node { $\beta_{8}$};
		\end{tikzpicture}
		\caption{Example partition of underside \(B_{U}\). Each actuator belongs to one partition \(\beta\) and its spatial characterization is assumed as zero outside its partition. Topside \(B_{T}\) can be partitioned similar to the presented one.}
		\label{fig:partition_input}
	\end{figure}
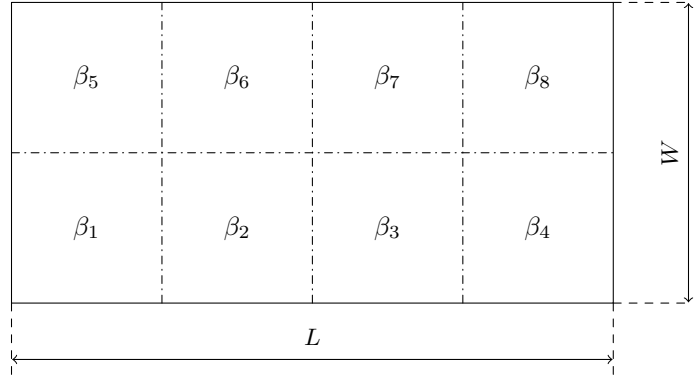

It is assumed that the underside \(B_{U}\) has a partition \(\beta_{n} \subset B_{U}\) with \( \bigcap\limits_{n=0}^{N_{u}-1} \beta_n = \left\{\right\} \) and \( \bigcup\limits_{n=0}^{N_{u}-1} \beta_n = B_{U} \) where \(N_{u}\) denotes the number of heating elements (see Figure \ref{fig:partition_input}). Each heating element has a spatial characterization \(b_{n}: B_{U} \rightarrow \left[0, 1\right]\) with restrictions
\begin{align*}
b_{n}(x) \geq&~ 0 \quad \text{for} ~ x \in \beta_{n} \text{,} \\
b_{n}(x) =&~ 0 \quad \text{for} ~ x \in B_{U} \setminus \beta_{n} 
\end{align*}
to describe the physical nature of the actuator including imperfection or abrasion. Each partition \( \beta_{j} \) correspond to the spatial characterization \(b_{j}\) and thus to its heating element and input signal \(u_{j}\). This means only separated inputs without any superposition are considered. Feedback design with Backstepping \cite{book:meurer2012control,book:krstic2008control} or optimal control \cite{book:hein2009mpc} can be found in the literature and will not be discussed here. The \(N_{u}\) actuators are assumed to have input signals \(u_{n}: \left[0,T_{final}\right) \rightarrow [u_{min}, u_{max}] \) with the bounds \(-\infty < u_{min} < u_{max} < \infty\). The induced heat flux is described by
\begin{align*}
\Phi_{in}(t,x) := b(x)^{\top} ~ u(t)
\end{align*}
for \( \left(t,x\right) \in \left[0, T_{final}\right) \times B_{U} \) and with 
\begin{align*}
b(x) = \left(b_{0}(x), b_{1}(x), \cdots, b_{N_{u}-1}(x) \right)^{\top}
\end{align*}
and
\begin{align*}
u(t) = \left(u_{0}(t), u_{1}(t), \cdots, u_{N_{u}-1}(t) \right)^{\top} \text{.}
\end{align*}

Analog to the plate's underside, it is assumed that the temperature on topside \(B_{T}\) is measured by \(N_{y}\) sensors where each of it correspond to a partition  \(\gamma_{n} \subset B_{T}\) with \(n \in \left\{0, 1, \cdots , N_{y}-1 \right\}\). The sensor's spatial characterization \(g_{n}: B_{T} \rightarrow \left[0,1\right] \) has to fulfill the requirements
\begin{align*}
g_{n}(x) \geq&~ 0 \quad \text{for} ~ x \in \gamma_{n} \text{,} \\
g_{n}(x) =&~ 0 \quad \text{for} ~ x \in B_{T} \setminus \gamma_{n} \text{.}
\end{align*} 
and additionally 
\begin{align*}
0 < \int_{\gamma_{n}} g_{n}(x) \opd x < \infty \text{.}
\end{align*}
The system output at the n-th sensor is defined by 
\begin{align}
y_{n}(t) := \frac{1}{\int_{\gamma_{n}} g_{n}(x) \opd x} ~ \int_{\gamma_{n}} g_{n}(x) ~ \heattx \opd x \text{.} \label{eq:output_sensor_measurement_NEW}
\end{align}
One notes, that the system output is bounded with \(0 \leq y_{n}(t) < \infty \) for all \(n \in \left\{0, 1, \cdots , N_{y}-1 \right\}\) and \( t \in \left(0, T_{final}\right] \).

As mentioned above, the choice of \(b_{n}\) and \(g_n\) depend strongly on the nature of the actuators and sensors. Here, the prototype formula 
\begin{align}
	b_{n}(x) = 
	\begin{cases}
	m \cdot \exp\left( - \lVert M (x - x_{c,n}) \rVert^{\nu} \right) \quad &\text{for} \quad x \in \beta_{n} \text{,} \\
	0 \quad &\text{for} \quad x \in B_{U} \setminus \beta_{n}
	\end{cases} \label{eq:spatial_char_prototype}
\end{align}
with \(m \in [0,1]\), \(M \in \mathbb{R}^{2 \times 2}\), \(\nu \geq 0\) and central point \(x_{c,n} \in \beta_{n} \) is chosen to describe the actuator's spatial characterization. The same formula \eqref{eq:spatial_char_prototype} is used to describe the sensor's spatial characterization \(g_{n}\) on \(B_{T}\). In the case of \(m = 1\) and \( M = 0_{2 \times 2} \) one holds the indicator function 
\begin{align*}
\tilde{b}_{n}(x) ~=~
\begin{cases}
1 \quad &\text{for} ~ x \in \beta_{n} \text{,} \\
0 \quad &\text{for} ~ x \in B_{U} \setminus \beta_{n} \text{.}
\end{cases}
\end{align*}

\section{Application to reduced two-dimensional case}
The heating plate is reduced to the two-dimensional side view \(\Omega = \left(0, L\right) \times \left(0, H\right)\) with \(L = 0.30\) and \(H = 0.01\) meter. The density \( \rho \) is assumed as constant, whereas the specific heat capacity and thermal conductivity are assumed as linear functions \(c(\temperature) =~ c_{0} + c_{1} \temperature\) and \(\lambda(\temperature) =~ \lambda_{0} + \lambda_{1} \temperature\), see Table \ref{table:phys_properties_const}. The heat equation in conservative form is noted as
\begin{align}
\int\limits_{\Omega} \rho ~ c(\heattx) ~ \frac{\partial \heattx}{\partial t} \opd x ~=~ \int\limits_{\Omega} \operatorname{div} \left[ \lambda(\heattx) ~ \nabla \heattx \right] \opd x \label{eq:heat_equation_conservative} 
\end{align}
with \( \heat(0,x) = \heat_{init}(x)\) for all \(x \in \Omega \) and boundary conditions as described by Equations \eqref{eq:bc_in_out} and \eqref{eq:bc_out_conduction_radiation} with constants as in Table \ref{table:phys_properties_const}. The initial temperature variation is assumed as \( \heat(0,x) = 300 + \delta(x) \) including a noise term 
\begin{align*}
\delta(x) = a_{0} ~  \cos \left( 2\pi ~ a_{1}  \frac{x_{1}}{L} \right) \cos \left(2 \pi ~ a_{2}  \frac{x_{2}}{H} \right)
\end{align*}
with \(a_{0} = 3\), \(a_{1} = 10\) and \(a_{2} = 5\). The number of heating elements and sensors is assumed as \(N_{u} = N_{y} = 5 \) and a naive proportional controller 
\begin{align*}
u_{n}(t) := 
\begin{cases}
K_{p,n} ~ e(t) \quad &\text{for} ~ e(t) > 0\\
0 \quad &\text{otherwise}
\end{cases}
\end{align*}
with \(e_{n}(t) := y_{ref} - y_{n}(t)\), \(K_{p,n} = 10^{4}\) for \(n \in \left\{0,1, \cdots, 4\right\} \) is designed to reach a constant reference temperature \(y_{ref} = 400\) Kelvin. The partitions of \(B_{U}\) and \(B_{T}\) are defined by
\begin{align*}
	\beta_{n} :=&~ \left(n ~ \frac{L}{N_{u}}, (n+1) ~ \frac{L}{N_{u}} \right) \times \left\{0\right\} \quad \text{and} \\
	\gamma_{n} :=&~ \left(n ~ \frac{L}{N_{y}}, (n+1) ~ \frac{L}{N_{y}} \right) \times \left\{L\right\} \text{.}
\end{align*}

Two test scenarios are considered for the actuators: the \textit{nominal} and the \textit{realistic} case with different values for the spatial characterization, see Table \ref{table:characterization_setting}. The sensors' spatial characterization is the same for both scenarios.

\begin{table}[b]
	\begin{center}
		\caption{\MakeUppercase{Spatial characterization for test scenarios}}
		\begin{tabular}{|l l l l|}
			\hline
			& m & M & \(\nu\) \\
			\hline
			Sensors & \(1.0\) & \(10.0\) & \(4.0\) \\[1ex]
			\hline
			Actuators (nominal) & \(1.0\) & \(0.0\) & \(4.0\) \\[1ex]
			Actuators (realistic) & \(1.0\) & \(30.0\) & \(4.0\) \\[1ex]
			\hline
		\end{tabular}
		\label{table:characterization_setting}
	\end{center}
\end{table}
\section{Spatial approximation}

The plate's volume \( \Omega \) is discretized with \(J \times K\) cells where \(J\) and \(K\) denote the number of cells in \(x_{1}\)- and \(x_{2}\)-direction, respectively. The finite volume method
\begin{align*}
p(t,x^{j,k}) = \frac{1}{\Delta x_{1} \Delta x_{2}} \int\limits_{x_{2}^{k-\frac{1}{2}}}^{x_{2}^{k+\frac{1}{2}}} \int\limits_{x_{1}^{j-\frac{1}{2}}}^{x_{1}^{j+\frac{1}{2}}} p(t,x) \opd x_{1} \opd x_{2}
\end{align*}
with \( \lvert \Omega_{j,k} \rvert = \Delta x_{1} \Delta x_{2} \) is used to approximate heat equation \eqref{eq:heat_equation_conservative} in each cell \(\Omega_{j,k}\) with \(j \in \left\{0, 1, \cdots, J-1\right\}\) and \(k \in \left\{0, 1, \cdots, K-1\right\}\). The left-hand side of the heat equation results in 
\begin{align*}
\frac{1}{\lvert \Omega_{j,k} \rvert} \int\limits_{\Omega_{j,k}} \rho ~ c(\heattx) ~ \dheattx \opd x \approx  \rho ~ c(\heat(t,x^{j,k})) ~ \dot{\heat}(t,x^{j,k}) \text{.}
\end{align*} 

The heat flux vector \(q(\heattx) = \lambda(\heattx) \nabla \heattx \) with its elements
\begin{align*}
q_{i}(\heattx) = \lambda(\heattx) \frac{\partial}{\partial x_{i}} \heattx
\end{align*}
are introduced to simplify the notation. Likewise, new (relative) coordinates 
\begin{align*}
\tilde{x} := x^{j,k} = 
\begin{pmatrix}
\left[j + \frac{1}{2}\right] \Delta x_{1} \\[1ex]
\left[k + \frac{1}{2}\right] \Delta x_{2}
\end{pmatrix}
\end{align*}
and \(\tilde{x} + \delta x_{1} = x^{j+1, k}\) with \(\delta x_{1} := \left( \Delta x_{1}, 0 \right)^{\top}\) are established. The right-hand side of Equation \eqref{eq:heat_equation_conservative} is separated for each flux as  
\begin{align*}
\frac{1}{\lvert \Omega_{j,k} \rvert} \int\limits_{\Omega_{j,k}} \operatorname{div}\left[ \lambda(\heattx) \nabla \heattx\right]  \opd x ~=&~ \frac{1}{\lvert \Omega_{j,k} \rvert} \int\limits_{\Omega_{j,k}} \operatorname{div} \left[q(\heattx)\right] \opd x \\
~=&~ \frac{1}{\lvert \Omega_{j,k} \rvert} \int\limits_{\Omega_{j,k}} \frac{\partial}{\partial x_{1}} q_{1}(\heattx) + \frac{\partial}{\partial x_{2}} q_{2}(\heattx) \opd x \text{.}
\end{align*}
From here on, only the flux \(q_{1}\) is considered - the calculations for \(q_{2}\) work similar. The integral over cell \( \Omega_{j,k}\) is solved and the derivatives \(\frac{\partial  q_{1}}{\partial x_{1}}\) and \(\frac{\partial  \heat}{\partial x_{1}}\) are approximated using finite differences. Consequently, one yields  
\begin{align}
	&\frac{1}{\lvert \Omega_{j,k} \rvert} \int\limits_{\Omega_{j,k}} \frac{\partial}{\partial x_{1}} q_{1}(\heattx) \opd x
	~=~  \frac{\Delta x_{2}}{\lvert \Omega_{j,k} \rvert} \left[ q_{1}(\heat\left(t, \tilde{x} + \frac{\delta x_{1}}{2}\right)) - q_{1}(\heat\left(t, \tilde{x} - \frac{\delta x_{1}}{2}\right)) \right] \nonumber \\
	&\quad ~=~ \frac{1}{\Delta x_{1}^2} \left[ \lambda(\heat\left(t, \tilde{x} + \delta x_{1}/2\right)) ~ \heat(t, \tilde{x} + \delta x_{1} ) + \lambda(\heat\left(t, \tilde{x} - \delta x_{1}/2\right)) ~ \heat(t, \tilde{x} - \delta x_{1} ) \right. \nonumber \\
	&\qquad \qquad \left. -  \left[\lambda(\heat\left(t, \tilde{x} + \delta x_{1}/2\right)) + \lambda(\heat\left(t, \tilde{x} - \delta x_{1}/2\right)) \right]  ~ \heat(t, \tilde{x} ) \right] \text{.} \label{eq:spatial_approx_dx1}
\end{align}
The thermal conductivity at the cell boundaries are approximated with
\begin{align*}
	\lambda(\heat\left(t, \tilde{x} + \delta x_{1}/2\right)) \approx \lambda( \left[\heat\left(t, \tilde{x}\right) + \heat\left(t, \tilde{x} + \delta x_{1} \right) \right]/2  ) \text{.}
\end{align*}

\begin{center}
	\begin{figure}
		\centering
		\begin{tikzpicture}[scale=0.5,line cap=round,line join=round,x=1.0cm,y=1.0cm]
		\draw (0.,0.) -- (20.,0.) -- (20., 3.) -- (0., 3.) -- cycle;
		\draw [dashed] (0.,-0.1) -- (0., -1.);
		\draw [dashed] (20.,-0.1) -- (20., -1.);
		\draw [dashed] (20.1, 0.) -- (21., 0.);
		\draw [dashed] (20.1, 3.) -- (21., 3.);
		\draw [<-, line width=1.2] (-2.5,1.5) -- (-0.1, 1.5);
		\draw [->, line width=1.2] (20.1,1.5) -- (22.5, 1.5);
		\draw [<-, line width=1.2] (10.,-2.5) -- (10.,-0.1);
		\draw [->, line width=1.2] (10.0,3.1) -- (10., 5.5);
		\draw[color=black] (0., -1.2) node {\small $x_{1}=0$};
		\draw[color=black] (20., -1.2) node {\small $x_{1}=L$};
		\draw[color=black] (22.7, 0.) node {\small $x_{2}=0$};
		\draw[color=black] (22.7, 3.) node {\small $x_{2}=H$};
		\draw[color=black] (-1.8, 2.7) node {$-\frac{\partial \heat}{\partial x_{1}}$};
		\draw[color=black] (24., 1.5) node {$\frac{\partial \heat}{\partial x_{1}}$};
		\draw[color=black] (8.3, -2.3) node {$-\frac{\partial \heat}{\partial x_{2}}$};
		\draw[color=black] (11.3, 5.2) node {$\frac{\partial \heat}{\partial x_{2}}$};

		\draw (0.,-7.5) -- (20.,-7.5) -- (20., -4.5) -- (0., -4.5) -- cycle;
		\draw [dotted] (0.,-7.5) -- (-2.,-7.5) -- (-2., -4.5) -- (0., -4.5);
		\draw [dotted] (20.,-7.5) -- (22.,-7.5) -- (22., -4.5) -- (20., -4.5);
		\draw [dotted] (0.,-7.5) -- (0.,-8.25) -- (20.,-8.25) -- (20.,-7.5);
		\draw [dotted] (0.,-4.5) -- (0.,-3.75) -- (20.,-3.75) -- (20.,-4.5);
		
		\foreach \k in {2, 4, 6,...,18}{				
			\draw [dashdotted] (\k, -3.75) -- (\k, -8.25);
		}
		\foreach \k in {0.75, 1.5, 2.25}{				
			\draw [dashdotted] (-2., -4.5 - \k ) -- (22., -4.5 -\k );	
		}
		
		\draw [dashed] (-2.,-7.9) -- (-2., -9.5);
		\draw [dashed] ( 0.,-8.5) -- ( 0., -9.5);
		\draw [dashed] ( 2.,-8.5) -- ( 2., -9.5);
		
		\draw [dashed] (18.,-8.5) -- (18., -9.5);
		\draw [dashed] (20.,-8.5) -- (20., -9.5);
		\draw [dashed] (22.,-7.9) -- (22., -9.5);
		
		\draw[color=black] (-2.3, -10.) node {\tiny $j=-\frac{3}{2}$};
		\draw[color=black] ( 0. , -10.) node {-\tiny $\frac{1}{2}$};
		\draw[color=black] ( 2. , -10.) node {\tiny $\frac{1}{2}$};
		\draw[color=black] (18. , -10.) node {\tiny $J-\frac{3}{2}$};
		\draw[color=black] (20. , -10.) node {\tiny $J-\frac{1}{2}$};
		\draw[color=black] (22. , -10.) node {\tiny $J+\frac{1}{2}$};

		\draw [dashed] (20.4, -8.25) -- (23.6, -8.25);
		\draw [dashed] (22.4, -7.5 ) -- (23.6, -7.5 );
		\draw [dashed] (22.4, -6.75) -- (23.6, -6.75);
		
		\draw [dashed] (22.4,-5.25) -- (23.6, -5.25);
		\draw [dashed] (22.4,-4.5 ) -- (23.6, -4.5);
		\draw [dashed] (20.4,-3.75) -- (23.6, -3.75);
		
		\draw[color=black] (25.0, -8.25) node {\tiny $k =-\frac{3}{2}$};
		\draw[color=black] (24.5, -7.5 ) node {\tiny $-\frac{1}{2}$};
		\draw[color=black] (24.2, -6.75) node {\tiny $\frac{1}{2}$};
		\draw[color=black] (24.7, -5.25) node {\tiny $K-\frac{3}{2}$};
		\draw[color=black] (25.2, -4.5) node {\tiny $K-\frac{1}{2}$};
		\draw[color=black] (24.7, -3.75) node {\tiny $K+\frac{1}{2}$};
		
		\end{tikzpicture}
		\caption{Two-dimensional heating plate with heat flux at boundaries (above) and Finite Volume grid (below).}
		\label{fig:boundary_cond_x1_x2}
	\end{figure}
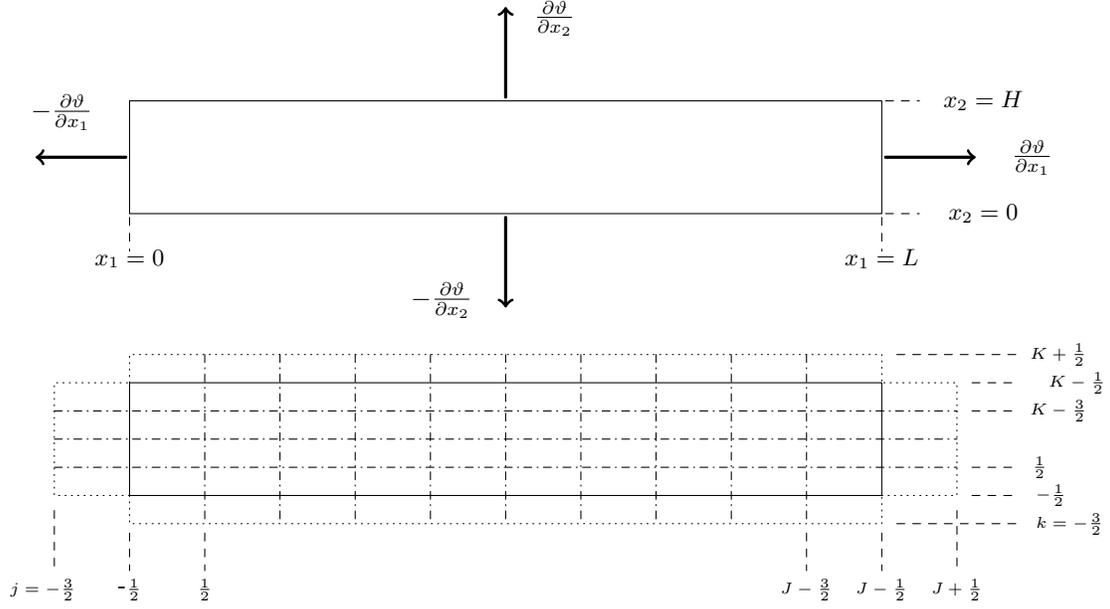
\end{center}

\subsection{Boundary conditions}
The boundary conditions are taken into account to complete the approximation as a high-dimensional ODE. Here, the cells at the left boundary with \(x_{1}=\frac{1}{2}\) and \(\tilde{x} = x^{0,k} \) are considered and therefore one holds 
\begin{align*}
\left.-\lambda(\heattx) \frac{\partial  \heattx}{\partial x} \right\rvert_{x= \begin{psmallmatrix}0 \\ x_{2} \end{psmallmatrix}} \approx \lambda( \heat(t, \tilde{x} - \delta x_{1}/2) ) \frac{1}{\Delta x_{1}} \left[ \heat(t, \tilde{x} - \delta x_{1}) - \heat(t, \tilde{x}) \right] = \phi_{out}(t, \tilde{x})
\end{align*}
and equivalently
\begin{align*}
\heat(t, \tilde{x} - \delta x_{1}) = \heat(t, \tilde{x}) + \frac{\Delta x_{1} ~ \phi_{out}(t, \tilde{x}) }{\lambda( \heat(t, \tilde{x} - \delta x_{1}/2) )} \text{.}
\end{align*}

The temperature of the \textit{virtual} cell \( \heat(t, \tilde{x} - \delta x_{1}) \) in Equation \eqref{eq:spatial_approx_dx1} is substituted with the boundary conditions to yield  

\begin{align*}
\frac{1}{\lvert \Omega_{0,k} \rvert} \int\limits_{\Omega_{0,k}} \frac{\partial}{\partial x_{1}} q_{1}(\heattx) \opd x
~=&~  \frac{1}{\Delta x_{1}^2} \left[ \lambda(\heat\left(t, \tilde{x} + \delta x_{1}/2\right)) ~ \heat(t, \tilde{x} + \delta x_{1} ) - \right. \\
& \qquad \left. \lambda(\heat\left(t, \tilde{x} + \delta x_{1}/2\right)) ~ \heat(t, \tilde{x}) \right] + \frac{1}{\Delta x_{1}} ~ \phi_{out}(t, \tilde{x}) \text{.}
\end{align*}
The same procedure is applied on the remaining boundaries with respect to the emitted and induced heat fluxes \(\phi_{out}\) and \(\phi_{in}\). The variables at the discrete points \(x^{j,k}\) are summarized as vectors and matrices
\begin{itemize}
	\item \(\Theta \in \mathbb{R}^{J \cdot K} \) for the temperature,
	\item \( C \in \mathbb{R}^{J \cdot K \times J \cdot K} \) for the specific heat capacity, 
	\item \( \Lambda_{1}, \Lambda_{2} \in \mathbb{R}^{J \cdot K \times J \cdot K} \) for the thermal conductivity and
	\item \(\Phi_{in}, \Phi_{out,x_{1}}, \Phi_{out,x_{2}} \in \mathbb{R}^{J \cdot K} \) for the induced and emitted heat flux.
\end{itemize}

The approximated heat dynamics is governed by the large-scale differential equation 

\begin{align*}
\rho ~ C(\Theta) ~ \dot{\Theta}(t) ~=~ \left[ \frac{1}{\Delta x_{1}^2}  \Lambda_{1}(\Theta(t)) + \frac{1}{\Delta x_{2}^2}  \Lambda_{2}(\Theta(t)) \right] \Theta(t) \\
+ \frac{1}{\Delta x_{1}} \Phi_{out,x_{1}}(t) + \frac{1}{\Delta x_{2}} \left[\Phi_{in}(t) + \Phi_{out,x_{2}}(t)\right] 
\end{align*}
with \( \Theta(0) = \Theta_{init}\).

\section{Simulation}

\begin{figure}[t]
	\centering
	\includegraphics[scale=0.38]{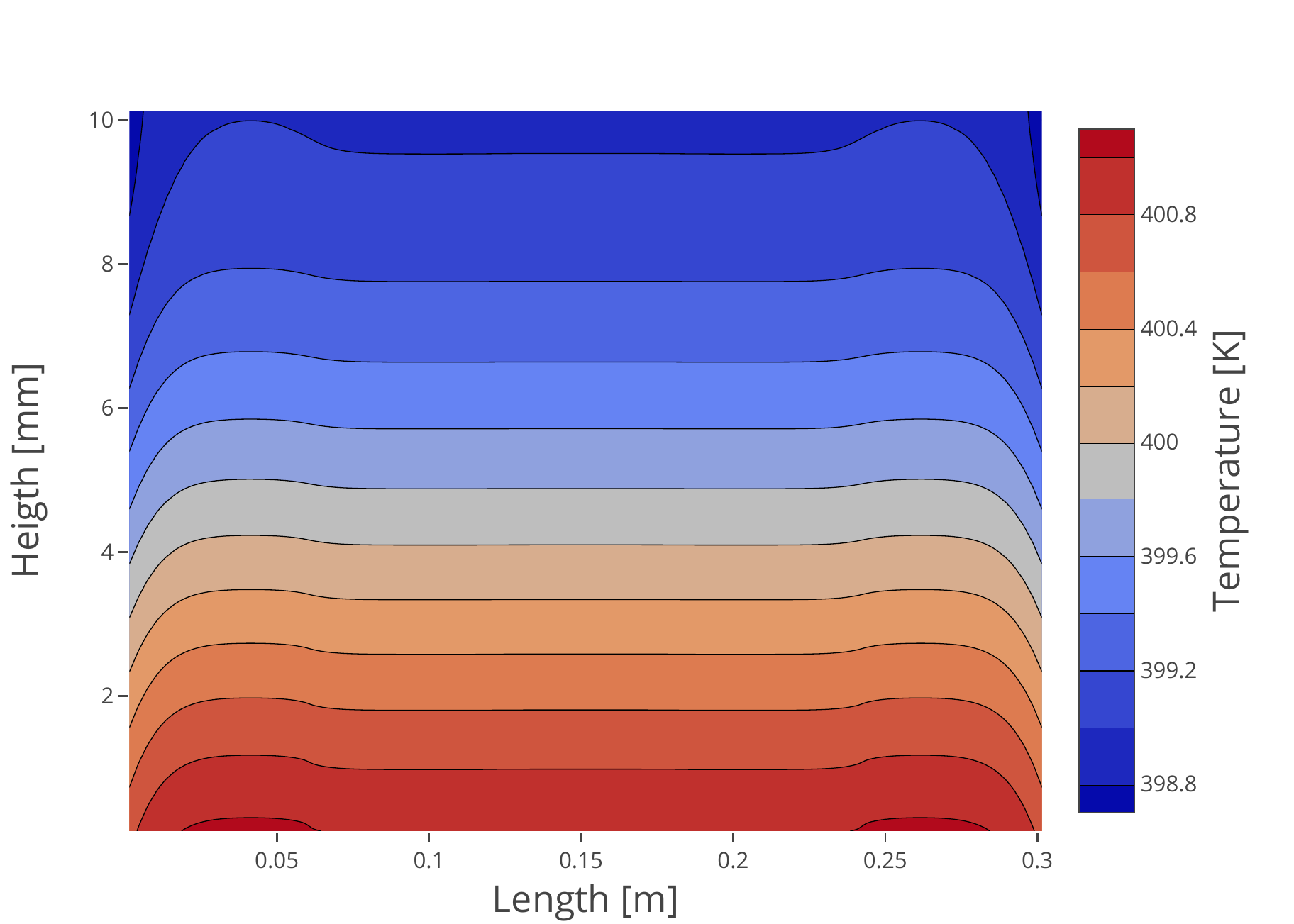} 
	\includegraphics[scale=0.38]{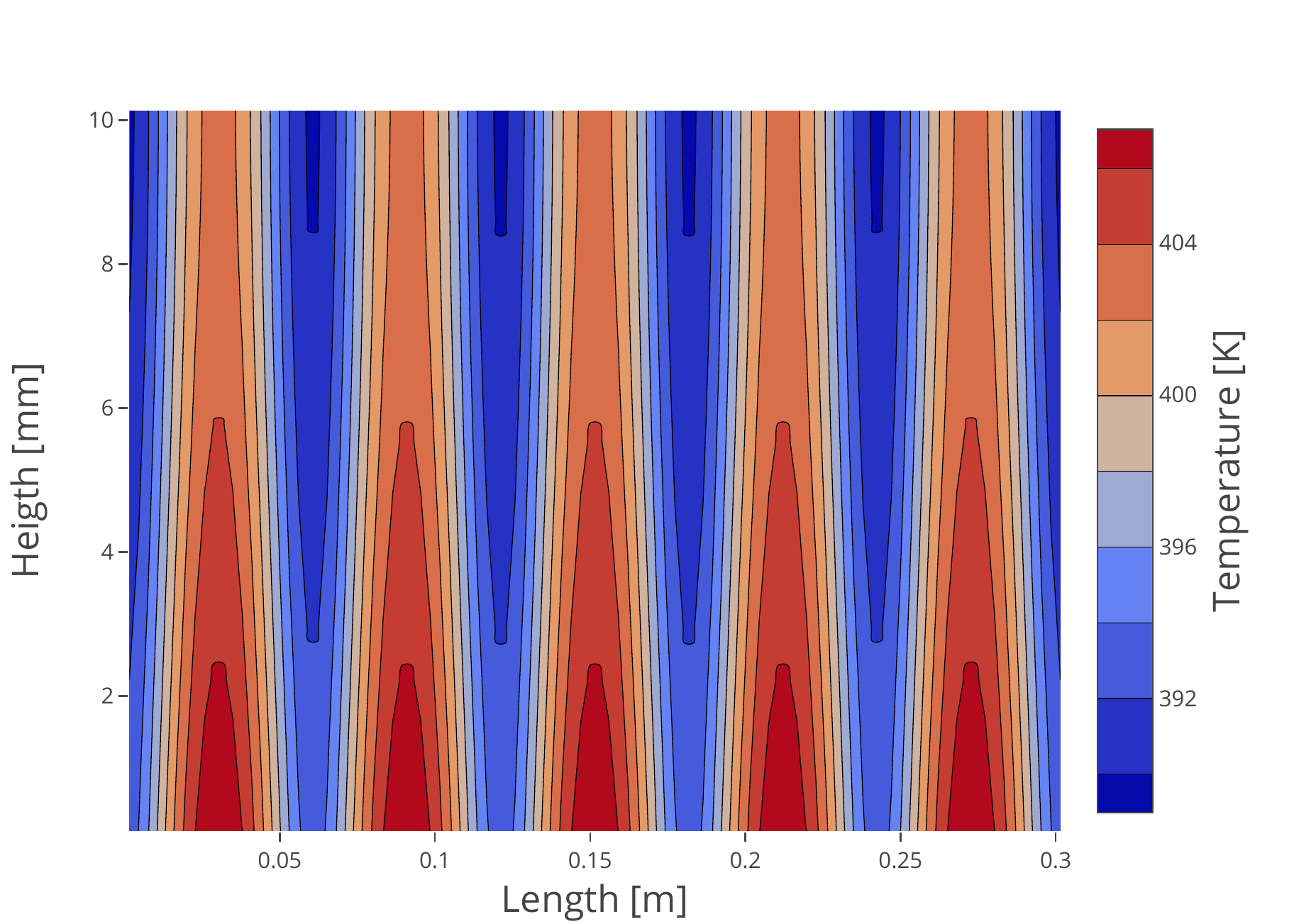} 
	\caption{Two-dimensional temperature distribution of Scenario 1 (left) and 2 (right). The comparison unveils a distinct difference of the thermal behavior in the whole plate between both scenarios.}
	\label{fig:temp_dist_2D}
\end{figure}

\begin{figure}[t]
	\centering
	\includegraphics[scale=0.38]{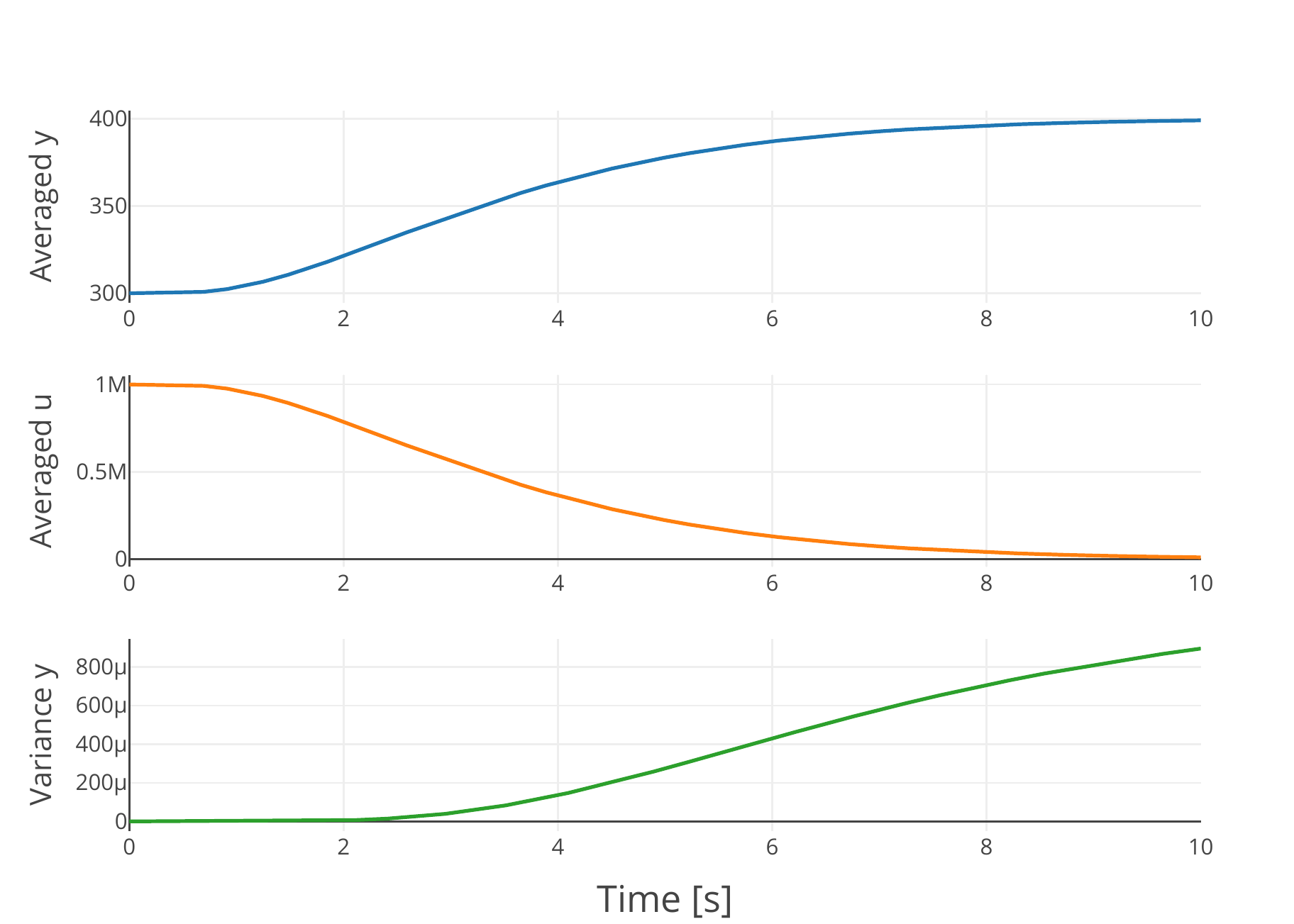}
	\includegraphics[scale=0.38]{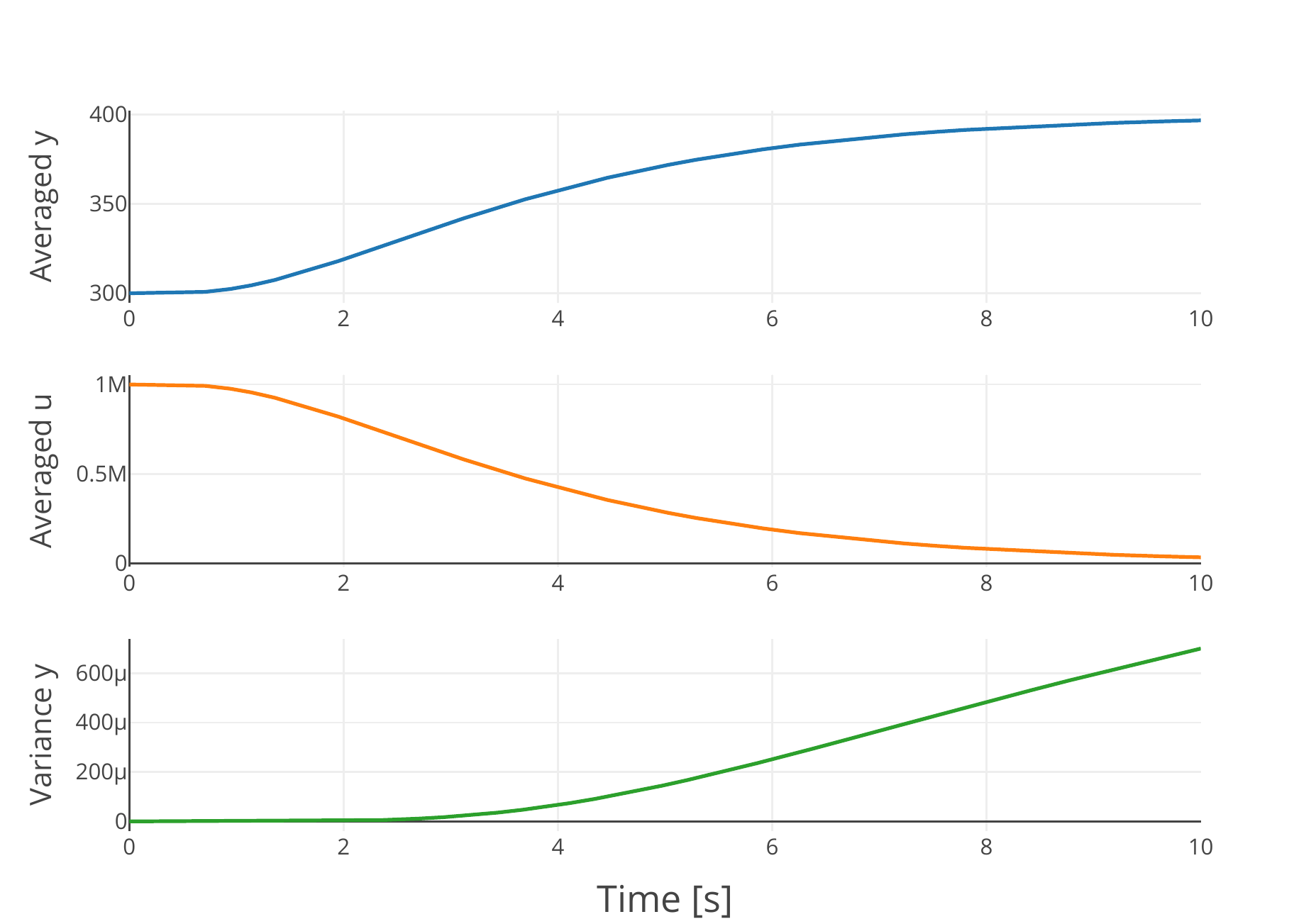}
	\caption{Averaged input and output signals of Scenario 1 (left) and 2 (right). The graphs show a comparable trend, even though the thermal behaviors are completely different in both cases.}
	\label{fig:averaged_input_output}
\end{figure}

The simulation is built with \textsc{Julia} \cite{article:bezanson2017julia} and the package \textsc{DifferentialEquations.jl} \cite{software:rackauckas2020sciml}. The visualization is carried out with \textsc{PlotlyJS.jl}. The source code of this simulation is available on GitHub:
\begin{center}
	\href{https://github.com/stephans3/MultipleSourceHeatingPlate2D.jl}{github.com/stephans3/MultipleSourceHeatingPlate2D.jl}.
\end{center}
The number of cells in horizontal and vertical axis  are set to \(J=100\) and \(K=40\) which corresponds to \(\Delta x_{1} = 3.0 \cdot 10^{-3}\) and \(\Delta x_{2} = 2.5 \cdot 10^{-4}\) meter, the final Time is set to \(T_{final} = 10\) and the sampling time is set to \(\Delta t = 10^{-3}\) seconds. The time integration was solved by the forward Euler method. 
\begin{figure}[h!]
	\centering
	\includegraphics[scale=0.60]{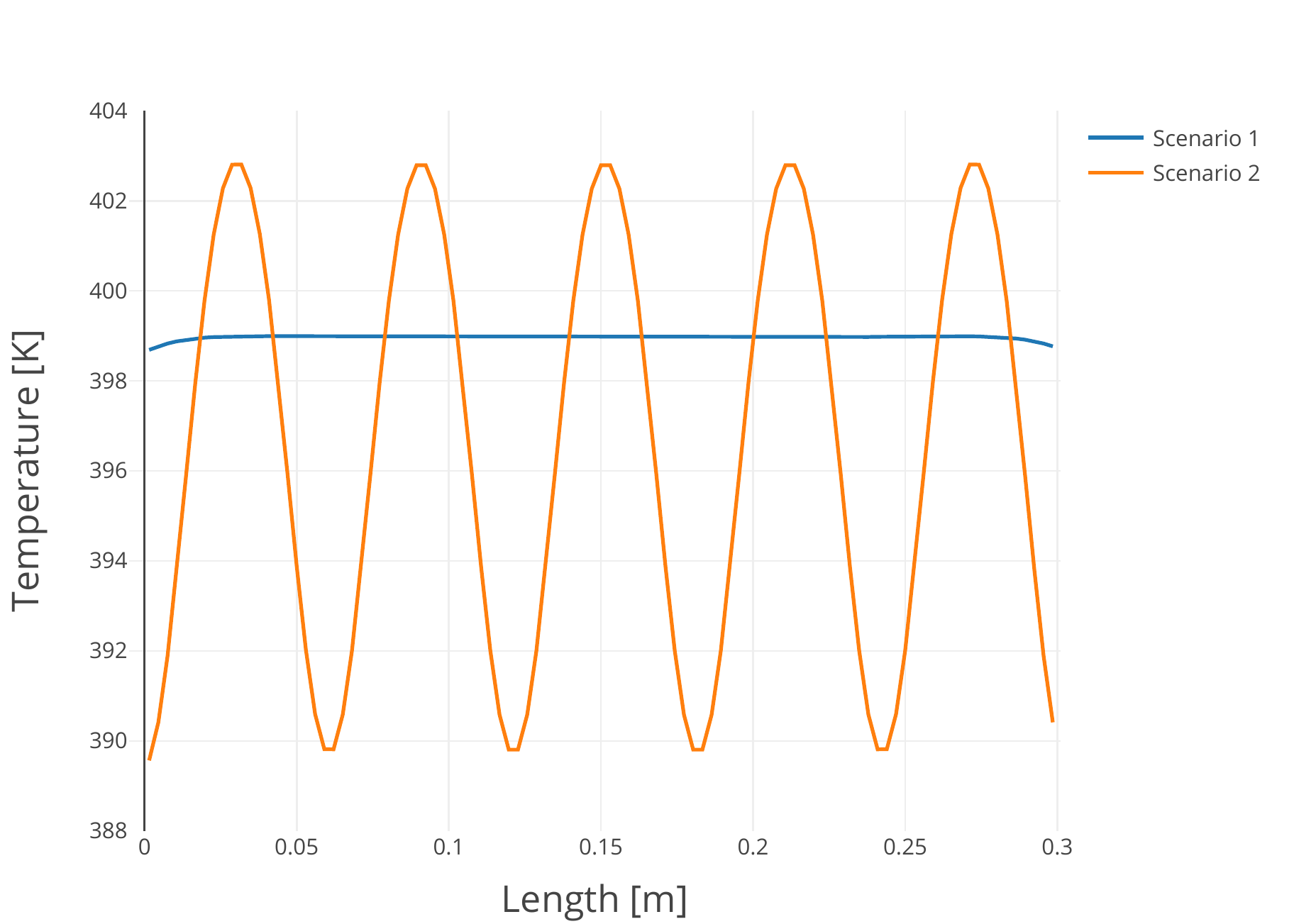}
	\caption{Temperature distribution on topside \(B_{T}\). The actuators' spatial characterization is transferred to the final temperature and can be observed in case of the nominal actuators (Scenario 1) as almost constant (desired), and in the case of realistic actuators (Scenario 2) as oscillating (undesired).}
	\label{fig:temp_dist_top}
\end{figure}  
The final temperature on topside \(B_{T}\), depicted in Figure \ref{fig:temp_dist_top}, shows an error from the reference \(y_{ref} = 400\) of approximately one Kelvin for the first scenario. However, in the second scenario the temperature distribution unveils an oscillatory behavior along the horizontal axis. This means, the \textit{realistic} actuators' spatial characterization cannot be compensated by the diffusive character of the heat equation. Same results can be observed in Figure \ref{fig:temp_dist_2D}: the temperature distributions - constant in Scenario 1 and oscillatory in Scenario 2 - are propagated through the two-dimensional plate. The naive controller approach leads to minimal differences of the input signals in both scenarios as presented in Figure \ref{fig:averaged_input_output}. The input and output averages are computed with
\begin{align*}
	\overline{u}(t) = \frac{1}{N_{u}} \sum_{n = 0}^{N_{u}-1} u_{n}(t) \quad \text{and} \quad \overline{y}(t) = \frac{1}{N_{y}} \sum_{n = 0}^{N_{y}-1} y_{n}(t) \text{.}
\end{align*}

Despite the comparable controller signals one has to note the distinct differences of the final temperature between the first and second scenario. Therefore, modeling systems like heating plates with multiple heat sources has to incorporate distributed parameter systems (up to three spatial dimensions), distributed actuators and sensors and their spatial characterization. Furthermore, a suitable controller has to be designed that can deal large-scale differential equations and the interaction between distributed actuators and sensors.

\section{Conclusion}

The thermal dynamics of a heating plate with multiple sources was discussed in the sense of a distributed parameter system in two (and three) dimensions. This general approach opens up the opportunity to analyze the complete heat evolution and the target temperature distribution in depth. The spatially distributed actuators and sensors were modeled including their typical characterization which has a significant influence on the heating process as illustrated in the numerical example. Although, the measured output reaches the reference temperature, the assumption of temperature-uniformity along the horizontal axis had to be rejected at least in the case of realistic actuators. This simple example points out the need to model thermal dynamics as precisely as possible despite its complexity and high computational effort. 
In future work the focus will lie on the development of a numerical framework to simulate and control two- and three-dimensional thermal processes with multiple heat sources on the underside and distributed sensors on the topside. This includes the modeling from a physical and analytical point of view, the spatial approximation with Finite Volumes and Galerkin approaches and the design of an efficient modern control scheme to compensate the (realistic) actuators' spatial character.
Finally, for verifying the examined thermal dynamic models, a heating plate set-up with multiple spatial symmetric Joule heating sources is available, which can be operated using either any single source or \(n \times n\) concurrent multiple sources with \(n\) up to five.

\section*{License}

This work is licensed under a Creative Commons \enquote{Attribution-ShareAlike 4.0 International} license.

\includegraphics[scale=1.]{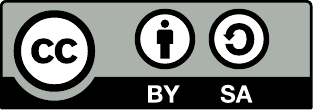}



\begin{thebibliography}{99}
	\bibitem{book:xiao2012introduction} Hong Xiao: \emph{Introduction to Semiconductor Manufacturing Technology}. Chapter 5: Thermal Processes. SPIE Press, Bellingham 2012.
	
	\bibitem{article:wei2019heat}
	Zhengying Wei, Jun Du: \emph{Heat and Mass Transfer of Additive Manufacturing Processes for Metals}. Heat and Mass Transfer-Advances in Science and Technology Applications. IntechOpen, 2019.
	
	\bibitem{article:zhou2011optimizing}
	Zhiwei Zhou, Boyan Song, Likuan Zhu, Zuntao Li, Yang Wang: \emph{Optimizing of Bioreactor Heat Supply and Material Feeding by Numerical Calculation}. International Conference on Intelligent Computing and Information Science. Springer, Berlin, Heidelberg, 2011.
	
	\bibitem{article:berger2007qualification}
	Lothar Berger, Peter Dress, Shun-Ho Yang, Chien-Hsien Kuo: \emph{Qualification of design-optimized multizone hotplate for 45-nm node mask making}. Photomask and Next-Generation Lithography Mask Technology XIV. Vol. 6607. International Society for Optics and Photonics, 2007.
	
	\bibitem{article:berger2004global}
	Lothar Berger, Peter Dress, Thomas M. Gairing, Chia-Jen Chen, Ren-Guey Hsieh, Hsin-Chang Lee, Hung-Chang Hsieh:
	\emph{Global critical dimension uniformity improvement for mask fabrication with negative-tone chemically amplified resists by zone-controlled postexposure bake}. Journal of Micro/Nanolithography, MEMS, and MOEMS 3.2 (2004): 203-211.
	
	\bibitem{article:tay2008predictive} A. Tay, K.K. Tan, S. Zhao, T.H. Lee: \emph{Predictive Ratio Control of Multizone Thermal Processing System in Lithography}. IFAC Proceedings Volumes 41.2 (2008): 10863-10868.
	
	\bibitem{book:han2009temperature} Yan Han: \emph{Temperature sensing and control in multi-zone semiconductor thermal processing}. Diss. 2009. Available: \url{https://scholarbank.nus.edu.sg/handle/10635/17322}.
	
	\bibitem{article:hamane2010thermal} Hiroto Hamane, Koudai Matuki, Fujio Hiroki, Kazuyoshi Miyazaki: \emph{Thermal MIMO controller for setpoint regulation and load disturbance rejection}. Control engineering practice 18.2 (2010): 198-208.
	
	\bibitem{article:joelianto2011bake} Endra Joelianto, Iqbal Ginanjar Prasetia: \emph{Bake Plate Control using A Robust Multiplexed Model Predictive Control (RMMPC)}. 2011 2nd International Conference on Instrumentation Control and Automation. IEEE, 2011.
	
	
	\bibitem{book:feng2011critical} Yong Feng: \emph{Critical dimension and temperature control in multi-zone thermal processing}. Diss. 2011. Available: \url{https://scholarbank.nus.edu.sg/handle/10635/35821}.
	
	\bibitem{article:jatunitanon2013multivariable} Poom Jatunitanon, Withit Chatlatanagulchai: \emph{Multivariable Robust Control for Two–Zone Thermal Plate System}. TSME International Conference on Mechanical Engineering (2013).
	
	
	\bibitem{article:jatunitanon2018robust} Poom Jatunitanon, Sarawoot Watechagit, Withit Chatlatanaguchai: \emph{Robust multi-model predictive control of multi-zone thermal plate system}. Songklanakarin Journal of Science \& Technology 40.1 (2018).
	
	\bibitem{article:krstic2005backstepping}
	Andrey Smyshlyaev, Miroslav Krstic: \emph{Backstepping observers for a class of parabolic PDEs}. Systems \& Control Letters 54.7 (2005): 613-625.
	
	\bibitem{article:krstic2015adaptive}
	T. Ahmed-Ali, F. Giri, M. Krstic, F. Lamnabhi-Lagarrigue, and L. Burlion: \emph{Adaptive Observer for a Class of Parabolic PDEs}. 	IEEE Transactions on Automatic Control 61.10 (2015): 3083-3090.
	
	\bibitem{article:kharkovskaia2020interval}
	Tatiana Kharkovskaia, Denis Efimov, Emilia Fridman, Andrey Polyakov, Jean-Pierre Richard: \emph{Interval observer design and control of uncertain non-homogeneous heat equations}. Automatica 111 (2020): 108595.
	
	
	
	\bibitem{article:eppler2001optimal}
	Karsten Eppler, Fredi Tröltzsch: \emph{Optimal Control Problems for the Nonlinear Heat Equation}
	Online Optimization of Large Scale Systems. Springer, Berlin, Heidelberg, 2001. 173-183.
	
	\bibitem{chapter:seidman1996control} Thomas Seidman: \emph{Control of the heat equation}. The Control Handbook. CRC Press, Boca Raton (1996): 1157-1168. Available: \url{http://www.math.umbc.edu/~seidman/Papers/ctrl_heq.pdf}.
	
	
	\bibitem{article:tan2000study}
	B.K. Tan, X.Y. Huang, T.N. Wong, K.T. Ooi: \emph{A study of multiple heat sources on a flat plate heat pipe using a point source approach}, International journal of heat and mass transfer 43.20 (2000): 3755-3764.
	
	\bibitem{article:song2016nonlinear}
	Yiming Song, Xiaoxiao Wang, Haipeng Teng, Yulei Guan: \emph{Nonlinear parametric predictive control for the temperature control of bench-scale batch reactor}. Applied Thermal Engineering, Volume 102 (2016): 134-143.
	
	\bibitem{article:tutcuoglu2016nonlinear}
	Abbas Tutcuoglu, Carmel Majidi, Wanliang Shan: \emph{Nonlinear thermal parameter estimation for embedded internal Joule heaters}. International Journal of Heat and Mass Transfer 97 (2016): 412-421.
	
	\bibitem{article:utz2011trajectory}  Tilman Utz, Thomas Meurer, Andreas Kugi: \emph{Trajectory planning for a two-dimensional quasi-linear parabolic PDE based on finite difference semi-discretizations}.  IFAC Proceedings Volumes 44.1 (2011): 12632-12637.
	
	\bibitem{article:xiao2016sliding} Tengfei Xiao, Han-XiongLi: \emph{Sliding mode control design for a rapid thermal processing system}. Chemical Engineering Science 143 (2016): 76-85.
	
	\bibitem{article:xiao2018dominant} Tengfei Xiao, Xiao-Dong Li, Shuqiang Wang: \emph{Dominant-Modes-Based Sliding-Mode Observer for Estimation of Temperature Distribution in Rapid Thermal Processing System}. IEEE Transactions on Industrial Informatics 15.5 (2018): 2673-2681.
	
	\bibitem{book:baehr2013heat} Hans Dieter Baehr, Karl Stephan: \emph{Heat and mass transfer}. Springer Science \& Business Media, Berlin 2013.

	\bibitem{book:meurer2012control} Thomas Meurer: \emph{Control of higher–dimensional PDEs: Flatness and backstepping designs}. Springer Science \& Business Media, Berlin 2012.

	\bibitem{book:krstic2008control} Miroslav Krstic, and Andrey Smyshlyaev: \emph{Boundary control of PDEs: A course on backstepping designs}. SIAM, Philadelphia 2008.
	
	\bibitem{book:hein2009mpc} Sabine Hein: \emph{MPC/LQG-Based Optimal Control of Nonlinear Parabolic PDEs}. Diss. 2009. Available: \url{https://nbn-resolving.org/urn:nbn:de:bsz:ch1-201000134}.
	
	
	\bibitem{article:bezanson2017julia} Jeff Bezanson, Alan Edelman, Stefan Karpinski, Viral B. Shah: \emph{Julia: A fresh approach to numerical computing}. SIAM Review 59.1 (2017): 65-98.
	
	\bibitem{software:rackauckas2020sciml} Christopher Rackauckas, et al.:\emph{SciML/DifferentialEquations.jl: v6.15.0}. [Computer software]. Zenodo (2020). Available: \url{https://doi.org/10.5281/ZENODO.3929125}.
	
	
	
	
	
	
	
		
\end{thebibliography}
\end{document}